\documentclass[conference]{IEEEtran}
\makeatletter
\newcommand{\linebreakand}{%
  \end{@IEEEauthorhalign}\hfill\mbox{}\par
  \mbox{}\hfill\begin{@IEEEauthorhalign}%
}
\makeatother

\usepackage{fancyhdr}                 %

\usepackage{amsmath,amssymb,amsfonts}
\usepackage{amsthm}
\usepackage{empheq}
\usepackage{algorithm}
\usepackage{algorithmic}

\usepackage{graphicx}
\usepackage[caption=false,font=footnotesize]{subfig}  %
\usepackage{tikz}

\usepackage{textcomp}
\usepackage{xcolor}
\usepackage{pifont}
\usepackage[inline]{enumitem}

\usepackage{booktabs}                 %
\usepackage{tabularx}
\usepackage{multirow}
\usepackage{makecell}                 %
\usepackage[flushleft]{threeparttable} %
\usepackage{etoolbox}
\makeatletter
\patchcmd{\@makecaption}{\normalfont\footnotesize\scshape #2}
  {\normalfont\footnotesize #2}{}
  {\errmessage{table caption patch did not apply to IEEEtran's \string\@makecaption}}
\makeatother
\usepackage{capt-of}                  %
\usepackage{colortbl}
\definecolor{zebra}{gray}{0.94}       %

\usepackage{siunitx}
\usepackage{mdframed}

\usepackage[hidelinks]{hyperref}      %
\usepackage{cleveref}

\hypersetup{
  pdftitle    = {FP64 Is All You Want, INT8 Is All You Need, FP4/6/8 Is All You Have},
  pdfauthor   = {Pratyai Mazumder, Alexandru Calotoiu, Torsten Hoefler},
  pdfkeywords = {Ozaki scheme, DGEMM emulation, matrix multiplication,
                 residue number system, combinatorial optimization,
                 low-precision arithmetic, Tensor Cores, GPU},
}

\usepackage{dblfloatfix}
\usepackage[defaultlines=2,all]{nowidow}
\usepackage[
  backend=biber,
  style=ieee,
  doi=true,
  url=true,
  sorting=none,                       %
  maxbibnames=6,                      %
  minbibnames=1,                      %
  giveninits=true,
  dashed=false,
]{biblatex}

\DeclareSourcemap{
  \maps[datatype=bibtex]{
    \map{
      \step[fieldsource=title,
            match=\regexp{\b(Adleman|Arm|Blackwell|Chinese|Cook|Gale|Instinct|Intel|Karatsuba|Mathemagix|Ozaki|Rivest|Shamir|Strassen|Toom|Server\s+Edition|Vera\s+Rubin|A100|H100|300X|355X|E2M1|V3|I(?=/))\b},
            replace=\regexp{\{$1\}}]
      \step[fieldsource=title,
            match=\regexp{\btoom-cook\b},
            replace=\regexp{\{Toom\}-\{Cook\}}]
      \step[fieldsource=title,
            match=\regexp{\bfp64-Emulated\b},
            replace=\regexp{\{FP64\}-emulated}]
    }
    \map[overwrite]{
      \step[fieldsource=entrykey, match=\regexp{\Auchino_high-performance_2025\Z}, final]
      \step[fieldset=booktitle, fieldvalue={Proceedings of the {SC} '25 Workshops of
            the International Conference for High Performance Computing, Networking,
            Storage and Analysis}]
    }
    \map[overwrite]{
      \step[fieldsource=entrykey, match=\regexp{\Abajard_generating_2021\Z}, final]
      \step[fieldset=booktitle, fieldvalue={2021 {IEEE} 28th Symposium on Computer
            Arithmetic ({ARITH})}]
    }
    \map[overwrite]{
      \step[fieldsource=entrykey, match=\regexp{\Abarrett_implementing_1987\Z}, final]
      \step[fieldset=booktitle, fieldvalue={Advances in Cryptology --- {CRYPTO} '86}]
    }
    \map[overwrite]{
      \step[fieldsource=entrykey, match=\regexp{\Adumas_finite_2002\Z}, final]
      \step[fieldset=booktitle, fieldvalue={Proceedings of the 2002 International
            Symposium on Symbolic and Algebraic Computation}]
    }
    \map[overwrite]{
      \step[fieldsource=entrykey, match=\regexp{\Abodrato_integer_2007\Z}, final]
      \step[fieldset=booktitle, fieldvalue={Proceedings of the 2007 International
            Symposium on Symbolic and Algebraic Computation}]
    }
    \map[overwrite]{
      \step[fieldsource=entrykey, match=\regexp{\Afp4-is-all-you-need_blas_2026\Z}, final]
      \step[fieldset=version, fieldvalue={commit 472b4c1}]
    }
  }
}

\AtEveryBibitem{%
  \iffieldundef{doi}{}{%
    \clearfield{url}%
    \clearfield{urlyear}%
    \clearfield{urlmonth}%
    \clearfield{urlday}%
    \clearfield{eprint}%
    \clearfield{eprinttype}%
    \clearfield{eprintclass}%
  }%
}
\AtEveryBibitem{%
  \iffieldundef{eprint}{}{%
    \iffieldequalstr{eprinttype}{arxiv}{%
      \clearfield{url}%
      \clearfield{urlyear}%
      \clearfield{urlmonth}%
      \clearfield{urlday}%
    }{}%
  }%
}
\AtEveryBibitem{\clearfield{note}}

\DeclareUnicodeCharacter{0394}{\ensuremath{\Delta}}
\DeclareUnicodeCharacter{03A3}{\ensuremath{\Sigma}}
\DeclareUnicodeCharacter{03BB}{\ensuremath{\lambda}}
\DeclareUnicodeCharacter{211D}{\ensuremath{\mathbb{R}}}
\DeclareUnicodeCharacter{2010}{-}
\DeclareUnicodeCharacter{2121}{TEL}
\DeclareUnicodeCharacter{FB02}{fl}
\usepackage[T1,OT1]{fontenc}
\DeclareTextAccentDefault{\k}{T1}

\mdfdefinestyle{calloutstyle}{%
  linewidth=0.6pt,
  roundcorner=2pt,
  innerleftmargin=6pt,
  innerrightmargin=6pt,
  innertopmargin=6pt,
  innerbottommargin=6pt,
  skipabove=6pt,
  skipbelow=6pt,
  backgroundcolor=black!3,
}
\newcommand{\rtc}{\ensuremath{R\mathbin{\triangleright}C}}
\newcommand{\ctr}{\ensuremath{C\mathbin{\triangleright}R}}

\definecolor{defgreen}{RGB}{0,110,50}
\newcommand{\defn}[1]{\textcolor{defgreen}{\textbf{#1}}}

\theoremstyle{plain}

\theoremstyle{definition}

\theoremstyle{remark}

\newtheorem{example}{Example}
\crefname{proposition}{Proposition}{Propositions}
\crefname{corollary}{Corollary}{Corollaries}

\newcommand{\searched}[1]{\,[\textup{S}\if\relax\detokenize{#1}\relax\else: #1\fi]}

\begin{document}

\title{FP64 Is All You Want, INT8 Is All You Need, FP4/6/8 Is All You Have
}

\author{%
\IEEEauthorblockN{Pratyai Mazumder}
\IEEEauthorblockA{\textit{ETH Zurich} \\
Zurich, Switzerland \\
\mbox{pmazumder@ethz.ch}}
\and
\IEEEauthorblockN{Alexandru Calotoiu}
\IEEEauthorblockA{\textit{ETH Zurich} \\
Zurich, Switzerland \\
\mbox{acalotoiu@ethz.ch}}
\and
\IEEEauthorblockN{Torsten Hoefler}
\IEEEauthorblockA{\textit{ETH Zurich} \\
Zurich, Switzerland \\
\mbox{htor@ethz.ch}}
}

\maketitle

\pagestyle{plain}

\thispagestyle{plain}

\begin{abstract}
Ozaki scheme II emulates FP64 matrix products with INT8 ones through residues
modulo pairwise coprime moduli, and variants for FP8 and FP4 have followed. We
treat these schemes as one family and pose the choice of a scheme as a
combinatorial program that minimizes the number of low-precision GEMMs.
Given, for each modulus, a finite set of ways to compute products modulo it
from low-precision GEMMs, we find the choice of moduli and ways with the
fewest GEMMs, for any format, accumulator and inner dimension, and derive
lower bounds on the GEMM count over the whole family. Applied to the
formats of current GPUs, the method gives the first FP6 schemes, an FP8 scheme
with fewer GEMMs than any previous one, and an FP4 scheme that the bounds show
needs the fewest GEMMs of any scheme in the family whose moduli lie in a
stated range. Implemented on three Blackwell GPUs, the INT8, FP8 and FP4
schemes run faster than native FP64, up to 83\texttimes{} on B300.
\end{abstract}


\section{Introduction}\label{sec:introduction}

\begin{figure}[t]
\centering
\includegraphics[width=\columnwidth]{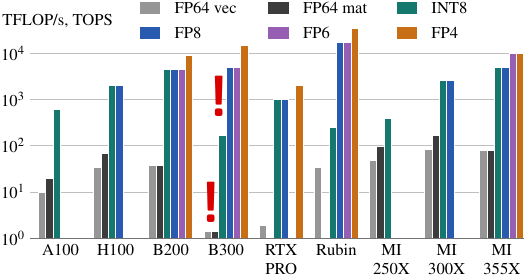}
\caption{Peak throughput per GPU, dense. FP64 alone is quoted
on both the vector lanes and the
matrix instruction. A blank slot is a format the GPU has no unit for, or a rate
its vendor does not publish. The two
``!'' mark B300's reduced FP64 and INT8 rates. Rubin's rates are
preliminary \cite{nvidia_nvidia_2026}, and NVIDIA gives its FP64 matrix rate
only by emulation \cite{aubrey_inside_2026}. RTX~PRO is the RTX~PRO~6000
Server Edition, for which NVIDIA publishes no FP64, INT8 or dense
low-precision rate; these are derived from its FP32 and its unlabeled FP8
and FP4 figures with the ratios NVIDIA gives for the chip
\cite{nvidia_rtx_pro_6000_server,nvidia_rtx_blackwell_pro_whitepaper}.}
\label{fig:hwgap}
\end{figure}

FP64 throughput has declined on recent NVIDIA and AMD GPUs, while peak
throughput for lower-precision FP formats has continued to
rise (\Cref{fig:hwgap}). Ozaki~I \cite{ozaki_error-free_2012}
and~II \cite{ozaki_ozaki_2025} compute an FP64 GEMM from many low-precision
ones, and were implemented on the much faster integer
units \cite{ootomo_dgemm_2024,uchino_performance_2025,uchino_high-performance_2025}. Those units stop at
eight bits: the MI355X datasheet lists INT4 at its INT8 rate \cite{amd_amd_2025},
Blackwell's lists no INT4 rate \cite{nvidia_b200_datasheet}, and its 4-bit
integer instruction runs at the 8-bit rate (\Cref{sec:boundary}). On B300
even the INT8 rate is reduced \cite{noauthor_nvidia_nodate-1}. On AI-focused GPUs
the fastest formats are sub-8-bit
floats \cite{nvidia_b200_datasheet,amd_amd_2025}.

Ozaki~I and~II have since been adapted to the
FP8 \cite{mukunoki_dgemm_2025,uchino_double-precision_2026,matsuoka_ozaki_2026} and
FP4 \cite{hayashi_dgemm_2026} units, where Ozaki~II needs fewer GEMMs than
Ozaki~I \cite{uchino_double-precision_2026,hayashi_dgemm_2026}. In FP8 it is
projected to run faster than native FP64 on B300 and
Rubin \cite{matsuoka_fp8_part1_2026,uchino_double-precision_2026,matsuoka_ozaki_2026}, and in FP4 it matches FP8 on an
RTX~PRO~6000 \cite{hayashi_dgemm_2026}. To our knowledge, only
particular instances of Ozaki~II for low-precision formats have been
described, each for one format.
The FP8 and FP4 ones split every residue into two low-precision values
\cite{uchino_double-precision_2026,hayashi_dgemm_2026}, the FP4 one at a fixed
base of $13$ and without Karatsuba's identity
\cite[\S1.3.2]{brent_modern_2010}. AWE \cite{hayashi_awe_2026} chooses the
base per modulus and multiplies integer combinations of such values, at $59$
GEMMs in FP4, selecting the moduli by an integer program whose formulation and
solver it does not give \cite[\S V-C]{hayashi_awe_2026}. No Ozaki~II scheme
has been given for FP6 or INT4; FP6 appears only in an Ozaki~I GEMM count
table \cite{mukunoki_dgemm_2025}.

Rather than another instance, we formulate the selection of an Ozaki~II
scheme as an optimization of its GEMM count on a broad class of formats and
accumulators. We choose jointly the moduli, the split of each residue and how
its product is formed. At a given inner dimension an exact search returns
a scheme of least GEMM count among its candidates, whose $K_{\max}$ is the
largest inner dimension its exactness conditions certify. Lower bounds on the
GEMM count over every way of forming a product show that at $K = 2^{14}$ no
FP4 scheme on a stated range of moduli needs fewer GEMMs than the one the
search returns. Individual formats follow as instantiations, and the INT8, FP8
and FP4 schemes the search chooses are timed on three Blackwell GPUs.

\subsection{Contributions}\label{sec:contributions}

\begin{enumerate}[leftmargin=1.3em,nosep]
  \item Ozaki~II scheme selection as a GEMM-count optimization for any
        format, accumulator and inner dimension, and an exact method for
        solving it over a given set of candidates. It gives the first FP6-based schemes,
        an FP8-based one of $30$ GEMMs against the previous best of $42$
        \cite{uchino_double-precision_2026}, and an FP4-based one of $59$, as
        does AWE \cite{hayashi_awe_2026}.
  \item Lower bounds on the GEMMs a modulus needs over every way of forming a
        product modulo it, and a certificate built from them: at $K = 2^{14}$, no
        FP4-based scheme on a stated range of moduli needs fewer than $59$
        GEMMs, and no INT8-based one fewer than $16$.
  \item GPU implementations of these schemes. On B300 the FP4-based
        schemes run up to $83\times$ faster than native FP64, and our
        implementation of GEMMul8's INT8 scheme runs there at least
        $1.08\times$ faster than GEMMul8 \cite{uchino_high-performance_2025}.
\end{enumerate}

\begin{figure*}[t]
\centering
\subfloat[$S$ itself.]{%
\label{fig:alphabet-a}%
\includegraphics[width=0.163\linewidth]{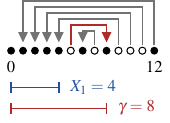}}%
\hfill
\subfloat[$S$ shifted by $13c_1$, and their union. Past the run, $58$, $60$ and
$64$ are reachable, while $57$, $59$, $61$ to $63$ and $65$ are not.]{%
\label{fig:alphabet-b}%
\includegraphics[width=0.817\linewidth]{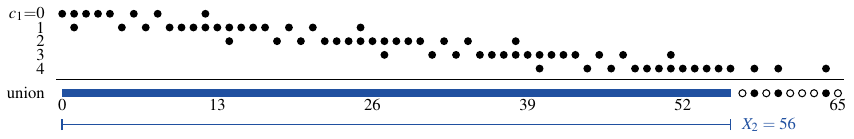}}
\caption{The reach of one limb against two, at base 13. Only the non-negative
half is drawn, E2M1 being symmetric; hollow marks are integers the set omits, and an
arrow points from a value to the least magnitude representing its class.
$X_\ell$ is the largest $X$ such that $\ell$ limbs represent all of
$[-X, X]$, and $\gamma$ the largest over the classes modulo $13$ of the least
magnitude in $S$ of the class.}
\label{fig:alphabet}
\end{figure*}

\section{Background}\label{sec:background}

\subsection{Emulation Problem}\label{sec:problem}

Given matrices $A$ of size $M \times K$ and $B$ of size $K \times N$ in FP64,
the closest FP64 result to $AB$ is $\mathrm{fl}(AB)$, with every entry of the
exact product rounded once. Emulation of $AB$ proceeds in three stages: convert
the operands to integers, form the integer product exactly on a low-precision
unit, and round once at the end. It returns $\mathrm{fl}(AB)$ if the
conversion is lossless.

Conversion scales by shared exponents. Let $\mu$ and $\nu$ have powers of two as
entries, one per row of $A$ and one per column of $B$, each the largest for
which every scaled entry of its row or column has magnitude below $2^{53}$
($1$ for a row or column of zeros), within the exact integer range of a 53-bit
significand \cite{ozaki_ozaki_2025,uchino_double-precision_2026}. With
$\operatorname{trunc}$ rounding toward zero elementwise, set
\begin{equation}
  A' = \operatorname{trunc}\bigl(\operatorname{diag}(\mu)\,A\bigr),
  \qquad
  B' = \operatorname{trunc}\bigl(B\,\operatorname{diag}(\nu)\bigr),
  \label{eq:convert}
\end{equation}
and recover the product as
$\operatorname{diag}(\mu)^{-1} (A'B') \operatorname{diag}(\nu)^{-1}$.

Scaling by a power of two is exact, so only $\operatorname{trunc}$ loses
information. An entry far below the largest in its row of $A$, or its column of
$B$, drops low-order bits. Every entry of $A'$ and $B'$ has magnitude below
$2^{53}$, so ${|a'_{ih} b'_{hj}| < 2^{106}}$ and
\begin{equation}
  \bigl| (A'B')_{ij} \bigr|
  \;=\; \Bigl| \textstyle\sum_{h=1}^{K} a'_{ih} b'_{hj} \Bigr|
  \;<\; K \cdot 2^{106} .
  \label{eq:range}
\end{equation}
No low-precision unit represents integers of that magnitude.

When the conditions of \Cref{sec:ov-exact} hold, $A'B'$ is formed exactly, and
the truncation in \eqref{eq:convert} and the final rounding are then the only
errors. Accuracy is then the same however the integer product is formed, so we
compare exact constructions by the number of low-precision GEMMs they issue.

\subsection{Ozaki II}\label{sec:ozaki}

Ozaki~I \cite{ozaki_error-free_2012} splits each operand into an unevaluated sum
of floating-point matrices whose pairwise products are exact. Ozaki~II
\cite{ozaki_ozaki_2025} instead reduces $A'$ and $B'$ modulo each of several
moduli $m$, taking representatives nearest zero. Since reduction preserves
products, one GEMM of the reduced matrices gives $(A'B') \bmod m$ after a final
reduction, provided their entries are values of the low-precision format and
the GEMM's sums are exact. The Chinese Remainder Theorem (CRT) then recovers
$A'B'$ from its residues modulo pairwise-coprime moduli whose product exceeds
the range of $A'B'$ \cite{garner_residue_1959}.

A single $k$-bit integer format can represent every residue of a modulus up to $2^k$.
On INT8, Ozaki~II therefore takes one GEMM per modulus (moduli up to $256$) and emulates
DGEMM with about $16$ GEMMs \cite{ozaki_ozaki_2025}.
The FP8 and FP4 formats hold only short runs of consecutive integers, so only
small moduli have all their residues in the format, and the product of such
moduli stays below the range of $A'B'$.
The FP-based schemes therefore write each reduced operand as $D_0 + bD_1$
with an integer $b$ and two low-precision matrices, which allows larger moduli at three or four
GEMMs per modulus \cite{uchino_double-precision_2026,hayashi_dgemm_2026}.

\section{A Family of Ozaki~II Schemes}\label{sec:overview}

Ozaki~II computes one full-precision GEMM from many low-precision ones, and
we minimize their number. This section defines the
schemes and their GEMM counts (\Cref{sec:ov-channels}), the forms built from
limbs (\Cref{sec:ov-cost}) and the constraints of exact emulation
(\Cref{sec:ov-exact}), poses the choice of a scheme as a combinatorial program
(\Cref{sec:ov-program}), solves it over candidate forms
(\Cref{sec:ov-solve}), and bounds its value over every form
(\Cref{sec:ov-lower}).

\subsection{Forms and Channels}\label{sec:ov-channels}

A format's \defn{alphabet} $S$ is its set of finite values scaled by the least
power of two $2^{e_S}$ that makes them all integers. At $e_S = 1$, E2M1 has an
alphabet of fifteen values with gaps,
$S = \{0, \pm1, \pm2, \pm3, \pm4, \pm6, \pm8, \pm12\}$. A low-precision GEMM
multiplies matrices whose entries lie in $S$. To obtain a product of integer
matrices modulo $m$ from such GEMMs, each entry is replaced by values in $S$
that depend only on its class modulo $m$, several GEMMs are issued, and their
results are combined with weights. A \defn{form of rank $r$ modulo $m$ over
$S$} is a family of tables $\alpha_k, \beta_k : \mathbb{Z}/m \to S$ and
integer weights $\lambda_k$, $k = 1, \dots, r$, such that for all
$t, u \in \mathbb{Z}/m$
\begin{equation}
  \sum_{k=1}^{r} \lambda_k\, \alpha_k(t)\, \beta_k(u) \;\equiv\; t\,u
  \pmod{m} .
  \label{eq:ov-form}
\end{equation}
Write $\operatorname{rank}(f)$ for the rank of a form $f$, and
\defn{$\operatorname{rank}_S(m)$} for the least rank of a form modulo $m$ over
$S$.

A form acts on the integer matrices $A'$ and $B'$ of \eqref{eq:convert}
entrywise. Its \defn{planes} $U_k = \alpha_k(A')$ and $V_k = \beta_k(B')$
apply the tables to the class modulo $m$ of every entry. The form issues the
$r$ GEMMs $W_k = U_k V_k$. Summing \eqref{eq:ov-form} over the $K$ terms of an
entry gives $\sum_k \lambda_k W_k \equiv A'B' \pmod{m}$ entrywise, which a
last pass, the \defn{epilogue}, computes. A form issues $r$ GEMMs whatever its
tables, so its cost is its rank (\Cref{fig:forms}). Where every class modulo $m$ has a member in
$S$, $S$ is \defn{complete} modulo $m$. One GEMM then suffices, with both
tables taking a member of the class and $\lambda_1 = 1$.

\begin{figure}[t]
\centering
\includegraphics[width=\columnwidth]{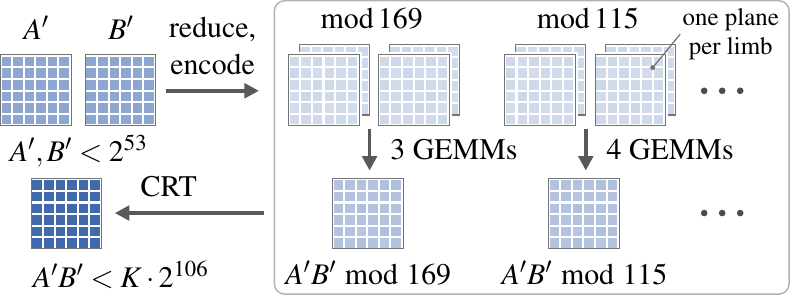}
\caption{A configuration at base $13$ over E2M1. Both operands are reduced onto
every channel and multiplied there as FP4 operand planes. CRT assembles the
residues. Two of nineteen coprime channels are drawn, one at
$m = b^2$ of rank three and one of rank four.}
\label{fig:channels}
\end{figure}

\begin{figure}[t]
\centering
\includegraphics[width=\columnwidth]{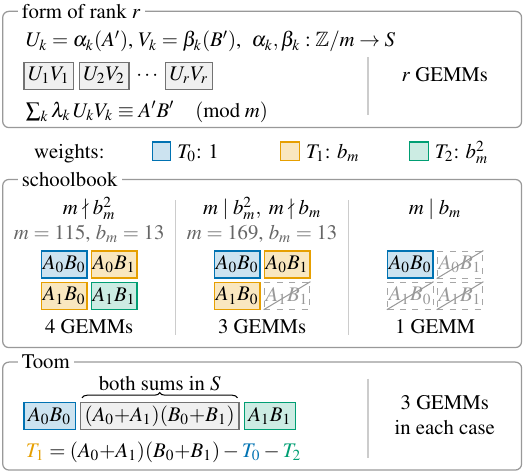}
\caption{GEMMs at a channel: a form of rank $r$, and the schoolbook and Toom
forms at two limbs in each case of $m$ against $b_m$ (instances from
\Cref{fig:channels}). Struck products lie in groups whose weight $m$
divides.}
\label{fig:forms}
\end{figure}

A \defn{channel} is a modulus $m$ with a form $f_m$ modulo $m$. A
\defn{configuration}, or scheme, is an alphabet with one channel at each modulus of a
pairwise-coprime set $\mathcal{M}$ (\Cref{fig:channels}). Its \defn{cost}
$Q = \sum_{m \in \mathcal{M}} \operatorname{rank}(f_m)$ is the number of
low-precision GEMMs it issues. It computes $A'B'$ modulo every $m \in \mathcal{M}$, hence modulo
$P_{\mathcal{M}} = \prod_{m \in \mathcal{M}} m$. The Chinese Remainder
Theorem (CRT) recovers $A'B'$ from those residues once $P_{\mathcal{M}}$ is
large enough (\Cref{sec:ov-exact}). The low-precision GEMMs dominate the run
time at large sizes (\Cref{sec:steps}), so the aim is the least $Q$, and at each modulus the
question is which forms of low rank exist over $S$.

\subsection{Forms Built from Limbs}\label{sec:ov-cost}

Forms of low rank come from two sources: the rank searches of
\Cref{sec:ov-lower}, which find them directly, and constructions that give a form
at every modulus they admit. The constructions here store each class as
limbs. At a \defn{base} $b \ge 2$, a tuple $c = (c_0, \dots, c_{\ell-1})$ of
\defn{limbs} $c_i \in S$ represents $\sum_i b^i c_i$, as digits do, except
that a limb may be negative and an integer may have several tuples
representing it. A modulus $m$ is \defn{admissible} at $\ell$ limbs and base
$b$ when every class modulo $m$ contains $\sum_i b^i c_i$ for some
$c \in S^{\ell}$.

A \defn{limb channel} $(m, b_m, \ell_m)$ is a channel whose form stores each
class modulo $m$ as $\ell_m$ limbs at base $b_m$; a one-limb channel needs no
base. The $i$th limbs of all entries of $A'$ form the \defn{limb plane}
$A_i$, and likewise $B_j$ of $B'$. The forms here are linear in the limbs:
each plane is a fixed integer combination of the limb planes,
$\sum_i p_{k,i} A_i$ on the A side and $\sum_j q_{k,j} B_j$ on the B side,
and with $w = (1, b_m, \dots, b_m^{\ell_m - 1})$ the weights satisfy
\begin{equation}
  \sum_{k=1}^{r} \lambda_k\, p_k\, q_k^{\top} \;\equiv\; w\, w^{\top}
  \pmod{m} ,
  \label{eq:ov-linear}
\end{equation}
so the products combine to the product of the represented values. Every
plane value is a GEMM operand, so the channel \defn{encodes} each class as a
tuple representing it whose plane values all lie in $S$, the one whose
largest plane value is least in magnitude. A schoolbook channel (below) takes
that least only over the tuples within its \defn{limb radii}
$\bar\rho_0, \dots, \bar\rho_{\ell_m-1}$, those with $|c_i| \le \bar\rho_i$
for every $i$. The channel's \defn{radius} $r_m$
is the largest plane-value magnitude over the tuples kept for its $m$
classes. Write
$S_\rho = \{\, c \in S : |c| \le \rho \,\}$. A channel whose limbs
lie within $r_m$ has $m \le |S_{r_m}|^{\ell_m}$.

\begin{example}\label{ex:ov-base}\label{ex:ov-radius}
At $b = 13$ over E2M1, $5 \notin S$ is represented as $-8 + 13 \cdot 1$. One
limb covers $[-4, 4]$ and two cover $[-56, 56]$ (\Cref{fig:alphabet}). At the
schoolbook channel $(115, 13, 2)$, class $57$ is stored as $(-6, -4)$, representing
$-58$. That channel's radius is $8$. With thirteen values within radius $8$,
two limbs admit no $m$ above $169$, which the channel $(169, 13, 2)$
attains. The ranges and the class $57$ modulo $115$ are the FP4 scheme's
\cite[\S III, \S V-A]{hayashi_dgemm_2026}.
\end{example}

The \defn{schoolbook} form takes every pair of limbs, $p = e_i$ and
$q = e_j$, the unit vectors of limbs $i$ and $j$, at weight $b_m^{i+j}$, since
$\sum_{i,j} b_m^{i+j} e_i e_j^{\top} = w w^{\top}$: it issues each cross
product $A_i B_j$ as a GEMM. Cross products of equal weight form a \defn{group}
$T_g = \sum_{i+j=g} A_i B_j$, which holds $n_g = \#\{(i, j) : i + j = g\}$ of
them. Once $m$
divides a weight $b_m^g$, it divides every heavier weight, so those groups
vanish modulo $m$ and are not formed. A schoolbook channel's form has rank
\begin{equation}
  q(m,b_m,\ell_m) \;=\; \ell_m^2 - \sum_{g \,:\, m \,\mid\, b_m^g} n_g ,
  \label{eq:qm}
\end{equation}
with the limbs as tables.

Write $A(x) = \sum_{i<\ell_m} A_i x^i$ and $B(x) = \sum_{j<\ell_m} B_j x^j$;
the groups $T_g$ are the coefficients of $A(x)\,B(x)$. That product has
degree $2\ell_m - 2$, so its values at $2\ell_m - 1$ distinct points
determine it. A \defn{Toom form} takes $p = q = (1, x_k, \dots,
x_k^{\ell_m - 1})$ at $2\ell_m - 2$ distinct integers $x_k$ and
$p = q = e_{\ell_m - 1}$ at the point $\infty$, a polynomial's value at
$\infty$ being its leading coefficient by convention
\cite[\S1.3.2]{brent_modern_2010}, \cite{bodrato_towards_2007}; so it
multiplies $A(x_k)$ by $B(x_k)$. At the points used here, $\{0, 1, \infty\}$
at two limbs and $\{0, 1, -1, 2, \infty\}$ at three, its weights are
integers, and it is a form of rank $2\ell_m - 1$. The FP8 Ozaki~II scheme
\cite[\S III-B]{uchino_double-precision_2026} and a multiple-precision INT8
one \cite[\S4.3]{kouya_ozaki_2026} take it at two limbs.

At three limbs the \defn{subset form} takes $p = q$ over the three unit
vectors and three of the four $0/1$ vectors with two or more ones
\cite{montgomery_five_2005,weimerskirch_generalizations_2006}: rank $6$,
against the Toom form's $5$, with plane values that are sums of at most three
limbs, where the Toom form's include $A_0 + 2A_1 + 4A_2$.

\subsection{Exactness Constraints and \texorpdfstring{$K_{\max}$}{Kmax}}\label{sec:ov-exact}

The limb and accumulator formats, the inner dimension $K$ and the forms
constrain how many channels are needed for exact emulation. Let every entry
of $A'$ have magnitude at most $B_A$ and every entry of $B'$ at most $B_B$.
CRT reconstructs an entry of $A'B'$ from its residues only up to multiples of
$P_{\mathcal{M}}$ \cite[\S 4.3.2, Thm.~C]{knuth_art_1997}, so $P_{\mathcal{M}}$ must exceed the width
$2 K B_A B_B$ of the range of $2 K B_A B_B + 1$ integers that the bounds allow,
the \defn{CRT window} \cite[eq.~(14)]{ozaki_ozaki_2025}:
\begin{equation}
  P_{\mathcal{M}} \;>\; 2 K B_A B_B .
  \label{eq:ov-crtgen}
\end{equation}
The FP64 operands of \eqref{eq:convert} have $B_A = B_B = 2^{53} - 1$, where
$2 K B_A B_B < K \cdot 2^{107}$ by \eqref{eq:range}, so
\begin{equation}
  P_{\mathcal{M}} > K \cdot 2^{107}
  \label{eq:ov-crt}
\end{equation}
suffices. Every number in this paper is for that window unless stated otherwise.
Call the computation that returns the result from the GEMM outputs the
\defn{decoder}. For $K \ge 3$, a decoder that reads the product only through its residues
modulo $P_{\mathcal{M}}$ and the scalings $\mu$, $\nu$ of \eqref{eq:convert},
as a configuration's does, needs a window within $\log_2(K/(K-2))$ bits of
\eqref{eq:ov-crtgen} even to return $\mathrm{fl}(A'B')$, or a faithful rounding
of $A'B'$, in place of $A'B'$.

A GEMM sums $K$ products of stored values. Write $\rho^A_k$ and $\rho^B_k$ for
the largest magnitudes plane $k$ stores on each side. The accumulator is exact
up to $\Lambda$, the largest integer such that the unit holds every integer in
$[-\Lambda, \Lambda]$, forms every product of two stored values lying there
exactly, and sums them without rounding: $2^{24}$ in FP32 and
$2^{31}-1$ in INT32. A channel adds its GEMMs into \defn{accumulators}, each
summing GEMMs of one weight, and which GEMMs share one is part of the channel.
A schoolbook channel is run \defn{per group}, each group in one accumulator,
or \defn{per GEMM}, each GEMM in its own; every other form here gives each
GEMM its own. An
accumulator that sums the GEMMs of a set $G$ of planes is exact where
\begin{equation}
  K \sum_{k \in G} \rho^A_k\, \rho^B_k \;\le\; \Lambda .
  \label{eq:ov-accgen}
\end{equation}
At a limb channel of radius $r_m$ whose accumulators sum at most $n_m$ GEMMs,
every term is at most $r_m^2$, so \eqref{eq:ov-accgen} holds at every
accumulator where
\begin{equation}
  n_m\, r_m^2\, K \;\le\; \Lambda .
  \label{eq:ov-acc}
\end{equation}
A schoolbook channel run per group has $n_m$ the size of its largest formed
group; run per GEMM it has $n_m = 1$, as every other form has. Since $n_m \le \ell_m$, a schoolbook
channel satisfies \eqref{eq:ov-acc}, whatever its base and grouping, whenever
its radius is at most
\begin{equation}
  r_{\max}(\ell_m) = \bigl\lfloor \sqrt{\Lambda / (\ell_m K)} \bigr\rfloor .
  \label{eq:rmax}
\end{equation}
A Toom channel satisfies it at radius up to $r_{\max}(1)$. With one product
per accumulator the bound is the delayed-reduction condition of
\cite[\S2.1.2]{dumas_dense_2008} and \cite[eqs.~(20)--(21)]{ozaki_ozaki_2025}.
The FP4 scheme and AWE state it with several products per accumulator
\cite[Lemma~2]{hayashi_dgemm_2026}, \cite[\S III-D]{hayashi_awe_2026}.
The bound \eqref{eq:ov-acc} is sufficient, not necessary: at a schoolbook
channel, \eqref{eq:ov-accgen} at an accumulator $G$ reads
$K \sum_{(i,j) \in G} \bar\rho_i \bar\rho_j \le \Lambda$ over its GEMMs
$A_i B_j$, so a limb that meets only smaller ones can exceed
$r_{\max}(n_m)$.

The epilogue combines the products in one of two orders.
\defn{Reduce-then-combine} (\rtc{}) reduces each product and each weight
modulo $m$ first. \defn{Combine-then-reduce} (\ctr{}) reduces once at the end
and takes each product unreduced. The epilogue's sums have a bound
$\Lambda_{\mathrm{ep}}$, defined as $\Lambda$ is; every configuration here
runs the epilogue in the accumulator's format, so
$\Lambda_{\mathrm{ep}} = \Lambda$. Write $G_1, \dots, G_s$ for the
accumulators of a channel, $\eta_j$ for the weight of the GEMMs of $G_j$ and
$\bar\eta_j$ for $\eta_j$ reduced into $[0, m)$. Under
\rtc{} every term is a product of two values in $[0, m)$. Under \ctr{} each
product enters unreduced. A product computed in slabs (\Cref{sec:implslabs})
adds two residues below $m$, which the term $2m$ allows for. The epilogue is
therefore exact where
\begin{subequations}\label{eq:ov-com}
\begin{empheq}[left=\empheqlbrace]{align}
  \max\Bigl\{2m,\ (m-1) \sum_j \bar\eta_j\Bigr\}
    &\;\le\; \Lambda_{\mathrm{ep}}, && \rtc,
  \label{eq:ov-comrtc}\\[4pt]
  \max\Bigl\{2m,\ K \sum_k |\lambda_k|\, \rho^A_k \rho^B_k\Bigr\}
    &\;\le\; \Lambda_{\mathrm{ep}}, && \ctr.
  \label{eq:ov-comctr}
\end{empheq}
\end{subequations}
Only the \ctr{} conditions involve $K$.

Solving \eqref{eq:ov-crtgen}, \eqref{eq:ov-accgen} and \eqref{eq:ov-comctr}
for $K$ gives the \defn{reconstruction cap}, \defn{accumulation cap} and
\defn{composition cap},
\begin{equation}
\begin{gathered}
  \kappa_{\mathrm{rec}} = \frac{P_{\mathcal{M}}-1}{2 B_A B_B}, \qquad
  \kappa_{\mathrm{acc}} = \min_{m \in \mathcal{M}}\ \min_{G}
    \frac{\Lambda}{\sum_{k \in G} \rho^A_k \rho^B_k}, \\[4pt]
  \kappa_{\mathrm{com}} = \min_{m \in \mathcal{M}}
    \frac{\Lambda_{\mathrm{ep}}}{\sum_k |\lambda_k|\, \rho^A_k \rho^B_k} ,
\end{gathered}
  \label{eq:caps}
\end{equation}
with $G$ over the accumulators of the channel at $m$. For the FP64 operands,
\eqref{eq:ov-crt} gives $\kappa_{\mathrm{rec}} = (P_{\mathcal{M}}-1)/2^{107}$.
A configuration's \defn{$K_{\max}$} is the largest $K$ at which these
conditions guarantee exactness:
$\min\{\lfloor\kappa_{\mathrm{rec}}\rfloor, \lfloor\kappa_{\mathrm{acc}}\rfloor\}$
under \rtc{}, provided \eqref{eq:ov-comrtc} holds, and
$\min\{\lfloor\kappa_{\mathrm{rec}}\rfloor, \lfloor\kappa_{\mathrm{acc}}\rfloor,
\lfloor\kappa_{\mathrm{com}}\rfloor\}$ under \ctr{}. Whichever cap attains
that minimum \defn{sets} $K_{\max}$. The FP4 scheme takes the same minimum at
its base-13 channels \cite[Lemma~3]{hayashi_dgemm_2026}. The window grows with $K$, so
$\kappa_{\mathrm{rec}}$ limits $K$ itself. The other two limit only how many
terms one accumulator sums. That count is $K$ unless a kernel splits the inner
dimension into slabs (\Cref{sec:implslabs}), which $K_{\max}$ assumes does not
happen.

\begin{figure}[t]
\centering
\includegraphics[width=\columnwidth]{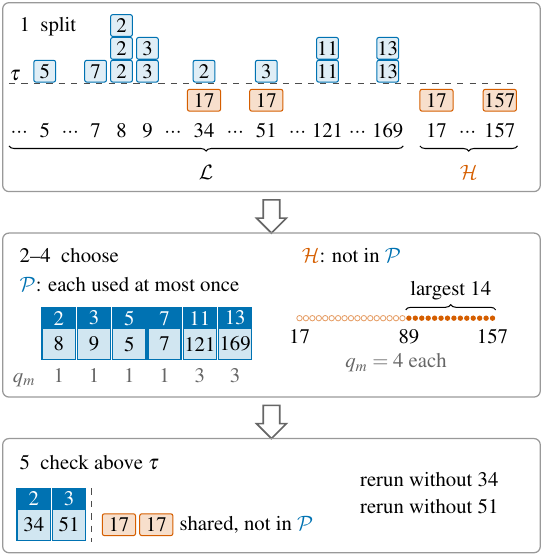}
\caption[The search on one pool]{The search on a
pool of E2M1's one-limb and two-limb schoolbook channels at $K = 2^{14}$ under
FP32 accumulation ($147$ channels, $\tau = 15$), panel by step. ($1$) Chips are each member's prime factors, blue up to $\tau$ and
vermilion above. ($2$--$4$) Each prime of $\mathcal{P}$ heads the one chosen member it
divides.
The set has $Q = 4 \cdot 1 + 2 \cdot 3 + 14 \cdot 4 = 66$, with
$P_{\mathcal{M}} = 2^{122.36} > K \cdot 2^{107}$. ($5$) $34$ and $51$ share
$17$, which is not in $\mathcal{P}$. The set found has no such pair.}
\label{fig:search}
\end{figure}

\subsection{Scheme Selection as a Combinatorial Program}\label{sec:ov-program}

Finding the best scheme can now be posed as one combinatorial program, to be
searched. Choosing a configuration at a given $K$ means choosing the moduli
and a form at each, so that $Q$ is least and the conditions above hold. Write
$\mathcal{F}_m(K)$ for the forms modulo $m$, each with its GEMMs partitioned
into accumulators of one weight, that satisfy
\eqref{eq:ov-accgen} at every accumulator and \eqref{eq:ov-comrtc} or
\eqref{eq:ov-comctr} for the epilogue order in use. The program is
\begin{equation}\label{eq:program}
\begin{aligned}
  Q^{\star} = \min_{\mathcal{M},\,\{f_m\}} \ & \sum_{m \in \mathcal{M}}
    \operatorname{rank}(f_m) \\
  \text{s.t.} \ & \gcd(m, m') = 1 \quad \forall\, m \ne m' \in \mathcal{M}, \\
  & P_{\mathcal{M}} > 2 K B_A B_B, \\
  & f_m \in \mathcal{F}_m(K) \quad \forall\, m \in \mathcal{M}.
\end{aligned}
\end{equation}

\subsection{Solving the Program}\label{sec:ov-solve}

Solving \eqref{eq:program} gives the best scheme for a format, but a search
over every form is intractable for wide formats. Fix at each modulus $m$ a
set $\mathcal{C}_m$ of \defn{candidate} forms. We give one tractable method
that solves \eqref{eq:program} exactly with every $f_m \in \mathcal{C}_m$,
and every configuration searched in this paper comes from it. Only coprimality and the window involve more than one channel.
Every other constraint concerns one channel, so each modulus can take its
form of least rank in $\mathcal{C}_m \cap \mathcal{F}_m(K)$. Write $q_m$ for that
rank, $\infty$ where it is empty. These forms, one at each
modulus of finite $q_m$, form the \defn{pool} $\mathcal{A}$. With every
$f_m \in \mathcal{C}_m$, \eqref{eq:program} reduces to
\begin{equation}
\begin{aligned}
  \min_{\mathcal{M} \subseteq \mathcal{A}} \ &
    \sum_{m \in \mathcal{M}} q_m \\
  \text{s.t.} \ & \gcd(m, m') = 1 \quad \forall\, m \ne m' \in \mathcal{M}, \\
                & P_{\mathcal{M}} > 2 K B_A B_B .
\end{aligned}
\label{eq:ov-reduced}
\end{equation}
The candidates here are the one-limb form; the schoolbook form at every base,
up to three limbs, every choice of limb radii and either grouping; the Toom
forms at every base and up to three limbs; the subset form; and over E2M1 the
rank-two, rank-five and rank-six forms the searches of \Cref{sec:ov-lower}
find, the \defn{searched forms}. The pools test a schoolbook
candidate by \eqref{eq:ov-accgen} at each accumulator and by
\eqref{eq:ov-comrtc} itself, a Toom or subset candidate by
\eqref{eq:ov-accgen} and by a bound that implies \eqref{eq:ov-comrtc}, and a
searched candidate by \eqref{eq:ov-accgen}, \eqref{eq:ov-comrtc} holding at
every searched form. At the
window \eqref{eq:ov-crt},
the search for \eqref{eq:ov-reduced} takes five steps (\Cref{fig:search}).
\begin{enumerate}[leftmargin=*,nosep]
  \item Split the pool $\mathcal{A}$ at a \defn{cutoff} $\tau$ such that each
    member has at most one prime factor above $\tau$, as RNS base generation
    splits its own candidates \cite{bajard_generating_2022}. The primes in
    $\mathcal{A}$ above $\tau$ form $\mathcal{H}$, the rest $\mathcal{L}$, and the
    primes up to $\tau$ that divide members of $\mathcal{L}$ form $\mathcal{P}$.
  \item For every total $Q_{\mathcal{H}}$, find the largest product
    $P_{\mathcal{H}}(Q_{\mathcal{H}})$ of members of $\mathcal{H}$ whose $q_m$
    sum to $Q_{\mathcal{H}}$.
  \item For every total $Q_{\mathcal{L}}$, find the largest product
    $P_{\mathcal{L}}(Q_{\mathcal{L}})$ of members of $\mathcal{L}$ whose $q_m$
    sum to $Q_{\mathcal{L}}$ and no two of which share a prime of
    $\mathcal{P}$. A dynamic program over the subsets of $\mathcal{P}$ does
    this, as a dynamic program over primes gives the largest coprime
    products in \cite[\S2]{deleglise_landaus_2008}.
  \item Choose $Q_{\mathcal{H}}$ and $Q_{\mathcal{L}}$ of least sum
    $Q = Q_{\mathcal{H}} + Q_{\mathcal{L}}$ such that
    ${P_{\mathcal{H}}(Q_{\mathcal{H}}) \cdot P_{\mathcal{L}}(Q_{\mathcal{L}})
    > K \cdot 2^{107}}$. The members in those two products form
    $\mathcal{M}$.
  \item If two members of $\mathcal{M}$ share a prime above $\tau$, run all
    five steps twice, once with each of the two removed from $\mathcal{A}$, and
    keep the result of lower $Q$.
\end{enumerate}
A configuration of least $Q$ at $K$
has the least $Q$ at every larger $K$ up to its $K_{\max}$. Searching again one past each
$K_{\max}$ found gives the least $Q$ at every larger $K$ (\Cref{fig:ksweep}).
Among the sets of least $Q$ the search keeps one of largest $K_{\max}$.
At a given $K$ and $\Lambda$, two alphabets agreeing on
$[-\Lambda/K,\, \Lambda/K]$ offer the same pool and the same least $Q$.

\begin{example}\label{ex:ov-equiv}
E4M3 agrees with E2M3 up to $60$, and E5M2 with E3M2 up to $448$, both below
$\Lambda/K = 1024$ under FP32 accumulation at $K = 2^{14}$, so that agreement
does not decide these pairs. Compared entry by entry, the pools of one- and
two-limb forms of each pair at that $K$ are the same, every value they
store lying within $r_{\max}(1) = 32$, so each FP8 alphabet takes
the configurations of its 6-bit \defn{partner} there.
\end{example}

\subsection{Certifying the Least GEMM Count}\label{sec:ov-lower}

The pool's least $Q$ is least over the candidates only. A form outside
them could have lower rank. Lower bounds on the rank decide this for a set of
moduli. Let $L(m) \le \operatorname{rank}_S(m)$ at every modulus of a set
$\mathcal{D}$. Then \eqref{eq:ov-reduced} over $\mathcal{D}$ with
$q_m = L(m)$ is at most $Q$ for every configuration on $\mathcal{D}$
that satisfies \eqref{eq:ov-crtgen}, whatever its forms, accumulators and
epilogue order. Where it equals the pool's least $Q$ and the pool's
configuration uses moduli of $\mathcal{D}$, this configuration is least over
every such configuration. The bounds are then a \defn{certificate} for
$\mathcal{D}$.

Over E2M1 at $K = 2^{14}$ a certificate holds for the moduli up to $474$
and the primes up to $246\,568\,393$: no configuration on them that
satisfies \eqref{eq:ov-crt} has $Q < 59$, and the configuration in
\Cref{sec:fp4} has $Q = 59$.

The bounds $L(m)$ come from the number of classes $S$ meets modulo $m$, from
finite searches at small ranks, from counts that bound the primes at which
each rank from seven to eighteen can occur, and at a composite from its prime
factors.
At a prime $p$, let $N_S(p)$ be the least $n$ for which some nonzero weights
$c_1, \dots, c_n$ make every class modulo $p$ a sum $\sum_j c_j s_j$ with
every $s_j \in S$. Every form modulo $p$ then has rank at least $2N_S(p) - 1$.
Over E2M1, $N_S(p) \ge 3$ at every prime above $225$, so every form modulo
such a prime has rank at least five.

One bound needs no search: every configuration that satisfies
\eqref{eq:ov-crtgen} has $Q \ge \log_2(2 K B_A B_B + 1) / \log_2 |S|$.

\begin{example}\label{ex:ov-int8}
INT8 has $256$ values, so at $K = 2^{14}$ every configuration that satisfies
\eqref{eq:ov-crt} has $Q \ge 16$, which the configuration of
\Cref{sec:boundary} attains.
\end{example}

Decoders other than a configuration's are bounded as well. A decoder that
combines all GEMM outputs by one weighted sum modulo an
$N > 2 K B_A B_B$ and returns $A'B'$ on every input splits into
channels. A certificate covers it where $N$ satisfies \eqref{eq:ov-crt},
those channels' moduli lie in the certificate's set and its tables send $0$
to $0$. Write $S \cdot S$ for the set of products of two members of
$S$. For $K \ge 3$ and the operands encoded entrywise, a decoder that
returns each entry of $A'B'$ from the $Q$ GEMM outputs at that entry and the
scalings of \eqref{eq:convert} has $|S \cdot S|^{Q}$ at least the number of distinct
products $ab$ with $|a| \le B_A$ and $|b| \le B_B$. One that returns
$\mathrm{fl}(A'B')$ has it at least the number of their roundings.

\begin{example}\label{ex:ov-decoders}
Over E2M1, $|S \cdot S| = 37$. More than $2^{99}$ products $ab$ are distinct,
and more than $2^{58}$ of their roundings, so a decoder that returns $A'B'$
needs $Q \ge 20$ and one that returns $\mathrm{fl}(A'B')$ needs $Q \ge 12$.
\end{example}

\section{Instantiation}\label{sec:instantiation}

We now apply the method of \Cref{sec:ov-solve} to find the configurations of
least $Q$ for the low-precision formats of current GPUs: E4M3, E2M3,
E2M1, INT4 and INT8.
Accumulation is FP32 for the
floats and INT32 for the integers, and the epilogue shares the accumulator's
format. \Cref{tab:moduli} gives their $Q$ at $K = 2^{12}$, $2^{14}$, $2^{16}$
and $2^{18}$, beside the prior schemes OzII-FP8 \cite{uchino_double-precision_2026},
OzII-FP4 \cite{hayashi_dgemm_2026}, AWE \cite{hayashi_awe_2026} and OzII-INT8
\cite{uchino_error_2026}.
\begin{figure}[t]
\centering
\includegraphics[width=\columnwidth]{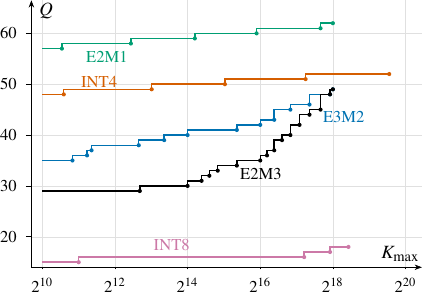}
\caption[Least $Q$ against the inner dimension]{Least $Q$ against the inner dimension $K$
over the pool of \Cref{sec:ov-solve}, each point at the largest $K$ its $Q$
suffices for, from $K = 2^{10}$. INT8's
level sets are one-limb and come from counting, not from the search.}
\label{fig:ksweep}
\end{figure}

\begin{table}[t]
\caption[The configuration each alphabet takes]{The least $Q$ each alphabet takes at four
inner dimensions over the pool of \Cref{sec:ov-solve}, by exhaustive search
or, at INT8, by counting (\Cref{ex:ov-int8}),
with the prior schemes OzII-FP8 \cite{uchino_double-precision_2026}, OzII-FP4
\cite{hayashi_dgemm_2026}, AWE \cite{hayashi_awe_2026} and OzII-INT8
\cite{uchino_error_2026}. Their rows are the published configurations, with a
dash where one is not exact at that $K$; over its own forms, with its slabs,
AWE's method reaches our $Q$ at every $K$ here. Every entry but AWE's, which
is its paper's, is exact under \rtc{}, and at $2^{14}$ the other float entries
under it only: \ctr{} caps OzII-FP8 at $K = 243$.}
\label{tab:moduli}
\centering
\begin{threeparttable}
\footnotesize
\setlength{\tabcolsep}{3pt}
\begin{tabular*}{\columnwidth}{@{\extracolsep{\fill}}lrrrr@{}}
\toprule
 & \multicolumn{4}{c@{}}{$Q$ at $K =$} \\
\cmidrule(l){2-5}
\textbf{$S$} & $2^{12}$ & $2^{14}$ & $2^{16}$ & $2^{18}$ \\
\midrule
OzII-FP8 \cite{uchino_double-precision_2026} & 42 & 42 & 42\tnote{$\|$} & --- \\
E4M3, E2M3 (ours) & 29 & 30 & 35 & 49 \\
\midrule
OzII-FP4 \cite{hayashi_dgemm_2026} & 75 & 75 & --- & --- \\
AWE \cite{hayashi_awe_2026} & 59 & 59 & --- & --- \\
E2M1 (ours) & 58 & 59\tnote{$\dagger$} & 61 & 62 \\
\midrule
INT4 (ours) & 49 & 50 & 51 & 52 \\
\midrule
OzII-INT8 \cite{uchino_error_2026} & 16 & 16 & 16 & --- \\
INT8 (ours) & 16 & 16\tnote{\P} & 16\tnote{\P} & 18 \\
\bottomrule
\end{tabular*}
\begin{tablenotes}\footnotesize
\item[$\dagger$] Also least over every configuration that satisfies
\eqref{eq:ov-crt} and whose moduli are all at most $474$ or primes up to
$246\,568\,393$, whatever its forms, accumulators and epilogue order
(\Cref{sec:ov-lower}).
\item[\P] Minimal over every configuration, with no search: the counting
bound of \Cref{ex:ov-int8} gives $16$ at $2^{16}$ as at $2^{14}$.
\item[$\|$] Exact at this $K$ with each square modulus keeping its three
products in separate accumulators, as GEMMul8 \cite{uchino_high-performance_2025}
runs this scheme; with its two cross products in one accumulator,
$K_{\max} = 32\,768$.
\end{tablenotes}
\end{threeparttable}
\end{table}

\Cref{fig:ksweep} plots the least $Q$ against $K$. From $2^{14}$ to
$2^{17}$ the 6-bit curves rise by up to $12$ GEMMs and the 4-bit ones by at
most $2$. At the 6-bit alphabets the accumulation cap limits $K$ at every
plotted point in that range: a larger $K$ lowers the sum over each
accumulator that \eqref{eq:ov-accgen} allows to $\Lambda/K$, so channels above
that bound leave the pool, and the coprime set of least $Q$ left satisfies
\eqref{eq:ov-crt} in more GEMMs. At INT4, and at E2M1 below $2^{18}$, the
reconstruction cap limits $K$, and each doubling of $K$ adds one bit to the
$P_{\mathcal{M}}$ that \eqref{eq:ov-crt} requires.

\subsection{FP8 and FP6: E4M3 and E2M3}\label{sec:fp8}\label{sec:fp6}

At $K = 2^{14}$ each FP8 alphabet takes its 6-bit partner's configurations
and GEMM counts (\Cref{ex:ov-equiv}), and E4M3 takes E2M3's at every $K$
searched above $3236$ (\Cref{fig:ksweep}), so we treat their schemes as one. Their
implementations differ in operand width and in the library that issues
their products (\Cref{sec:gemms}). E5M2 and E3M2 are not pursued. Within
radius $32$ they hold $33$ values against E4M3's and E2M3's $49$, so they need
smaller moduli and more of them.

At $K = 2^{14}$ the E4M3 configuration of least $Q$ takes $30$ GEMMs: $15$
one-limb channels at one GEMM each and $5$ two-limb channels at three. By contrast, OzII-FP8 puts every channel at two limbs and three
GEMMs, with no one-limb channels: $36$ at the $12$ moduli its own condition
$P_{\mathcal{M}} > 2^{107}$ needs, which omits $K$, and $42$ at the $14$
moduli that \eqref{eq:ov-crt} needs at this $K$.

\subsection{FP4: E2M1}\label{sec:fp4}

OzII-FP4 fixes base $13$ with every
channel at two limbs, $75$ GEMMs. Each choice of \Cref{sec:overview} lowers
that count in turn. Base $11$ needs one channel fewer, every channel still
two-limb, for $71$ GEMMs. Mixing one- and two-limb channels at base $11$
brings it to $68$. A
base per channel brings the count to $66$, and Toom forms to $59$, the count
AWE \cite{hayashi_awe_2026} reaches with the same moduli
(\Cref{tab:moduli}).

\subsection{INT4 and INT8}\label{sec:boundary}

INT4 is $\{-8,\dots,7\}$, the whole 4-bit two's-complement range.
On B200, B300 and RTX~PRO~6000 no instruction multiplies 4-bit integers
faster than 8-bit ones: the warp-level 4-bit MMA is issued as two 8-bit
integer MMAs of half the depth, and the \texttt{tcgen05} MMA of B200 and B300
has no 4-bit integer kind \cite{nvidia_parallel_nodate}. INT4 configurations
issue far more GEMMs than INT8. So, of the integer
alphabets, only INT8's configurations have been evaluated for Blackwell GPUs.

INT8 is $\{-128,\dots,127\}$. Under INT32 accumulation at $K = 2^{14}$,
$r_{\max}(1) = 362$ lies above every value it holds, so every modulus up to
$256$ is admissible at one limb and costs one GEMM. Sixteen
pairwise-coprime ones satisfy \eqref{eq:ov-crt}, and no INT8 configuration
takes fewer (\Cref{ex:ov-int8}). Of the sets of sixteen, ours keeps its
moduli up to $239$, whose radius of $119$ gives the largest $K_{\max}$,
$151\,647$; OzII-INT8's reach $256$ and stop at $131\,071$.

\section{Implementation}\label{sec:implementation}

Our implementation is a CUDA library that runs a configuration as
\Cref{sec:ov-channels} defines it: at each channel, the tables $\alpha_k$ and
$\beta_k$ of its form, its weights $\lambda_k$ and the accumulators its GEMMs
add into. It takes any form as its tables, a searched form among them, and
builds the tables itself for the one-limb, schoolbook, Toom and subset forms
from a modulus, a base and a limb count.
Building a
configuration checks \eqref{eq:ov-accgen} at every accumulator and a bound
that implies \eqref{eq:ov-comrtc} at every channel, from the tables and
weights the kernels receive, and refuses a $K$ above $\kappa_{\mathrm{rec}}$.
One build runs every configuration of \Cref{tab:moduli,fig:ksweep}, the prior
schemes' included, but INT4's and AWE's. Every size here and in \Cref{sec:evaluation} is a square
product, $M = N = K = n$.

\subsection{Structure}\label{sec:structure}

Computing one product $C = AB$ through our implementation, a \emph{call}, follows
the path of \Cref{fig:channels} in five phases:
\begin{enumerate}[leftmargin=1.3em,nosep]
  \item Scale finds one exponent per row of $A$ and column of $B$.
  \item Reduce applies Scale's exponents and truncates each entry, giving
        $A'$ and $B'$ without storing them. For each channel it reduces them
        modulo $m$ and writes each distinct plane $\alpha_k(A')$ and
        $\beta_k(B')$ of the channel's form once.
  \item The GEMMs $W_k = U_k V_k$ add into one array per accumulator.
  \item The epilogue computes each channel's $\sum_k \lambda_k W_k \bmod m$
        from its accumulators.
  \item Reconstruct applies the CRT, rounds once and unscales.
\end{enumerate}
Only the GEMMs run on the low-precision unit; the other four phases are custom
vector CUDA kernels.
Every GEMM tile that needs a plane reads it rather than forming it again. The channels run one after another, so the workspace holds every channel's
planes and residues but only one channel's accumulator arrays.

\subsection{Slabs}\label{sec:implslabs}

Past a configuration's $K_{\max}$, either the scheme changes or, where
the accumulation cap set that $K_{\max}$, the inner dimension breaks into
\emph{slabs} of $K_{\mathrm{t}}$ terms, each accumulated on its own and added
into a running residue. The \emph{blocked} call takes $K_{\mathrm{t}}$ from
the caller, and serves a $K$ whose planes do not fit in memory as well; each
slab runs its own epilogue over the whole output.

\subsection{The GEMMs}\label{sec:gemms}\label{sec:blockscale}\label{sec:traversal}

E4M3 and E2M1 products go through cuBLASLt, INT8's through CUTLASS on B200 and
cuBLASLt on B300 and RTX~PRO~6000. The 6-bit formats have no cuBLASLt type, so
CUTLASS issues them. On B300 that path runs B200's dense kernel, which is
exact on B300 as well.

\paragraph{Block scales}
The dense instruction takes a 4-bit operand at the 8-bit rate and the
block-scaled one reaches the 4-bit rate, so E2M1 runs on the block-scaled one.
Every block scale is $1$, which its scale formats hold exactly, so no product
changes and \eqref{eq:ov-accgen} bounds every partial sum in whatever order the
unit adds.

\paragraph{Tile order}
The products CUTLASS issues take output tiles in bands of columns,
\emph{banding}, where the scheduler's own order takes whole columns and at
$n = 2^{14}$ streams more than L2 holds between two reads of a plane's strip.
We set the band width per GPU and separately for INT8 through CUTLASS's
swizzle parameter \cite{nvidia_cutlass_2026}; cuBLASLt takes its own order.
Banding shortens B200's INT8 call at $8$ tiles wide.

\subsection{Around the GEMMs}\label{sec:aroundgemms}

\paragraph{Reduce}
Reduce reads each class's plane values from its form's table; an INT8 channel
stores the balanced residue in two's complement and reads none. A modulus
inside one byte reduces in two $32$-bit Barrett steps
\cite{barrett_implementing_1987} rather than one with a $128$-bit multiply.

\paragraph{Epilogue}
The epilogue reduces each accumulator before combining, the order \rtc{},
which every float configuration needs to be guaranteed exact
(\Cref{tab:moduli}). Each accumulator enters at its
weight reduced into $[0, m)$, whatever the form, so one kernel computes the
weighted sum at every form, in the accumulator's format.

\paragraph{Reconstruct}
Reconstruct forms the CRT sum in digits whose sums stay exact, takes the
balanced representative from one quotient estimate with an exact fallback,
and rounds once.

\section{Evaluation}\label{sec:evaluation}

\begin{figure*}[t]
\centering
\includegraphics[width=\textwidth]{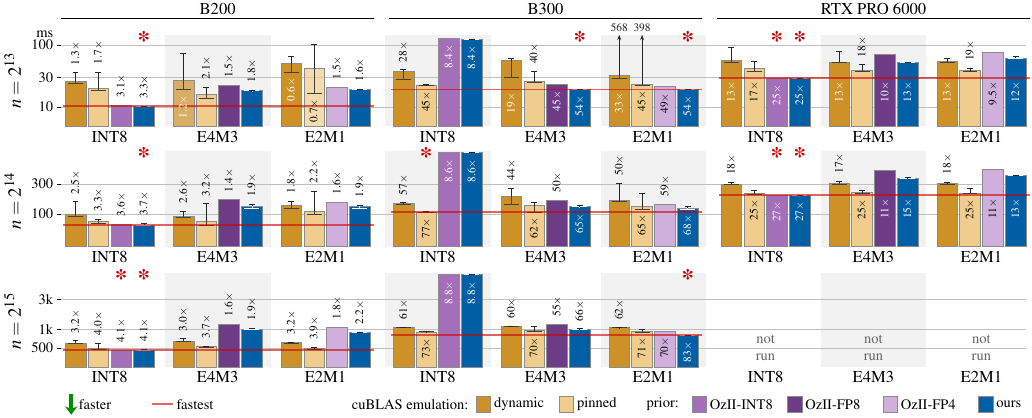}
\caption[Time of one call]{Median
time of one call, against the vendor's emulation in both its modes
\cite{nvidia_cublas_2026,schwarz_guaranteed_2026}, the prior scheme of the
group's alphabet ---
OzII-INT8 \cite{uchino_error_2026} on INT8, OzII-FP8
\cite{uchino_double-precision_2026} on E4M3, OzII-FP4 \cite{hayashi_dgemm_2026}
on E2M1 --- and ours. A label is that bar's speedup over
native FP64, which is not drawn; each row's axis is logarithmic over decades
of its own. A whisker spans the lowest and highest call behind the median,
two of them leaving the row with the figure they reach. Each group is one
process, so each cuBLAS mode is timed once in each configuration's process,
three times a GPU and size; a prior bar is our time scaled by that scheme's
median ratio from another harness. A red line marks the
fastest bar of each GPU at each size, and a red asterisk each bar whose label
reads the same as that bar's.}
\label{fig:times}
\end{figure*}

Ours, native and the vendor's emulation run in one process per GPU and
configuration, interleaved, and no call is counted until its result has been
checked against native (\Cref{tab:setup}). That pairing makes every ratio among
them a comparison of calls made alternately in one process, not of separate
runs. The prior schemes'
bars in \Cref{fig:times} and the released implementations of \Cref{fig:prior}
come from separate processes.

\begin{figure*}[t]
\centering
\includegraphics[width=\textwidth]{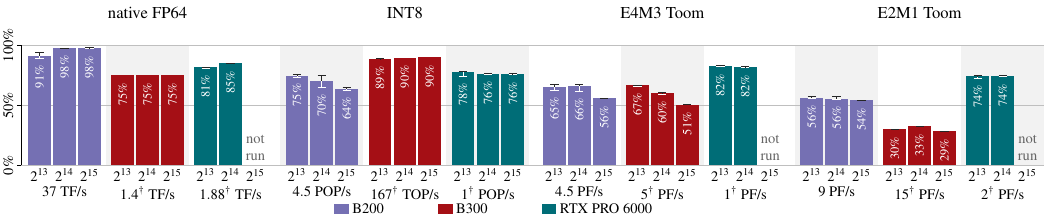}
\caption[Share of the datasheet peak reached]{The share of its format's
datasheet peak each method reaches, by GPU and size; for ours the bar is the
median of the four profiled calls and the whisker their lowest and highest,
and for native the median and extremes of the three unprofiled processes of
\Cref{fig:times}. Under each
group is the peak taken as $100\%$, with $\dagger$ on one derived from a rack
figure or a published ratio rather than given per GPU. Ours is profiled under CUDA~13.0.2 with cuBLAS~13.1, not
the toolkit of \Cref{tab:setup}, and with the E4M3 and E2M1 products through
CUTLASS rather than cuBLASLt. Native is the vendor's own kernel against its
own peak.}
\label{fig:peakshare}
\end{figure*}

\begin{figure*}[t]
\centering
\includegraphics[width=\textwidth]{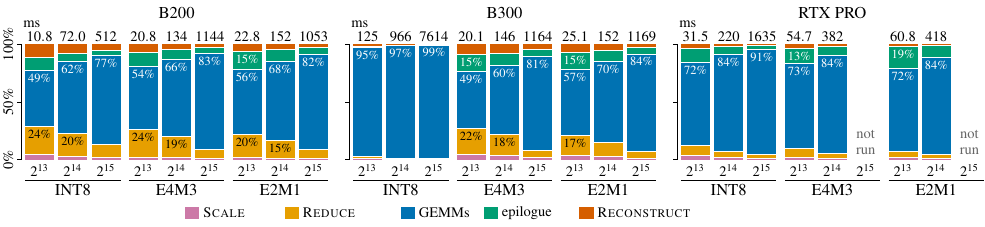}
\caption[Where a call's time goes]{Where one call's time goes, by
GPU, configuration and size; E4M3 and E2M1 take the Toom form. Each phase's
median is taken separately, so the five need not sum to the call and the
shares are of their sum. Profiled under CUDA~13.0.2 with cuBLAS~13.1, not the toolkit of
\Cref{tab:setup}, and with the E4M3 and E2M1 products through CUTLASS rather
than cuBLASLt.}
\label{fig:phaseshare}
\end{figure*}

\begin{table}[t]
\caption[Measurement setup]{GPUs, software, methods
and protocol of the run-time measurements.}
\label{tab:setup}
\centering
\footnotesize
\begin{tabularx}{\columnwidth}{@{}l@{\hspace{7pt}}X@{}}
\toprule
\multicolumn{2}{@{}c@{}}{\textit{Hardware}} \\
GPUs & B200 \cite{nvidia_b200_datasheet} (\texttt{sm\_100}, $178$~GiB); B300
  SXM6 \cite{noauthor_nvidia_nodate-1} (\texttt{sm\_103}, $268$~GiB); RTX~PRO~6000 Blackwell Server Edition (\texttt{sm\_120},
  $95$~GiB). Cloud containers with one GPU each. \\
\midrule
\multicolumn{2}{@{}c@{}}{\textit{Software}} \\
Driver & 580.95.05. \\
Toolkits & CUDA 13.4.1 with cuBLAS 13.8.0.4, installed over the one PyTorch
  ships and preloaded; the Python harnesses map it alone. \\
Kernels & CUTLASS 4.7.0 \cite{nvidia_cutlass_2026} and cuBLASLt: E4M3 and
  E2M1 products through cuBLASLt, INT8's through CUTLASS on B200 and
  cuBLASLt on B300 and RTX~PRO~6000. \\
\midrule
\multicolumn{2}{@{}c@{}}{\textit{Baselines}} \\
Native & \texttt{cublasDgemm}. \\
cuBLAS & Emulated DGEMM \cite{nvidia_cublas_2026,schwarz_guaranteed_2026},
  Ozaki~I in fixed point;
  \texttt{PERFORMANT}: dynamic precision, and mantissa pinned at $55$ bits. \\
OzII-FP8 & The FP8 scheme \cite{uchino_double-precision_2026} ($Q = 42$), run
  as OzII-FP4 is. \\
OzII-FP4 & The base-13 E2M1 scheme \cite{hayashi_dgemm_2026}
  ($Q = 75$), run by our implementation beside our own configurations in one
  process,
  $16$ rounds, on plain $\mathcal{N}(0,1)$ operands. \\
OzII-INT8 & The INT8 modulus set of \cite{uchino_error_2026} ($Q = 16$,
  moduli up to $256$ where ours stop at $239$), run the same way, $8$ rounds;
  its product is ours bitwise. \\
\midrule
\multicolumn{2}{@{}c@{}}{\textit{Ours}} \\
Configs & INT8 ($Q = 16$), E4M3 Toom ($30$; $34$ at $32\,768$), block-scaled
  E2M1 Toom ($59$; $60$ at $32\,768$): at each $n$ the least $Q$ with
  $K_{\max} \ge n$ (\Cref{fig:ksweep}). \\
\midrule
\multicolumn{2}{@{}c@{}}{\textit{Protocol}} \\
Operands & Square, $n \in \{8192, 16\,384, 32\,768\}$; row $i$ of each operand
  is $\mathcal{N}(0,1)$ times $2^{e_i}$, $e_i$ an integer drawn from
  $[-20, 20)$ separately for $A$ and $B$. \\
Timing & Methods interleaved, $4$ to $16$ repetitions, one process per GPU
  and configuration, each timed call after its own untimed one. \\
Speedup & Baseline time over ours, per repetition. \\
Profiles & Synchronized between phases; speedups are unprofiled. \\
Checks & Timed only within $10^{-12}$ (ours) or $10^{-6}$ (other emulations)
  of native in relative Frobenius norm. \\
\bottomrule
\end{tabularx}
\end{table}

\subsection{Speedup}\label{sec:speedup}

Every configuration is faster than native FP64
wherever both were timed (\Cref{fig:times}). The INT8 configuration is at
least $3.1$ times slower than either cuBLAS mode on B300, where its rate is
cut to $167$ TOP/s. Against that emulation the float
configurations are faster on B300 at every size in its dynamic mode, and in
its pinned one everywhere but E4M3 at $2^{15}$, each against the calls of its
own process; across the three processes the pinned mode reads $108$ to
$134$~ms at $2^{14}$, and at $108$~ms it is the fastest method there. On B200
and the RTX~PRO~6000 the pinned mode is the faster of the two, except for
E2M1 at $2^{13}$ on B200, where the pinned calls of that process were slower
than in fifteen others. The E2M1 configuration is faster than OzII-FP4, and
the E4M3 configuration than OzII-FP8, at every size timed on the three GPUs. OzII-INT8's time over the INT8
configuration's lies between $0.96$ and $1.06$ at every size but $2^{13}$ on
B200, where it is $1.04$ to $1.09$.
The speedup over native FP64 rises
with $n$ in every configuration on every GPU.

\subsection{Throughput}\label{sec:throughput}

Native FP64 reaches three quarters of its own peak or better, and our GEMMs,
with the E4M3 and E2M1 products through CUTLASS, at least half of the
datasheet peak of the format they use, E2M1 on B300 aside
(\Cref{fig:peakshare}). That one exception is under a third, against an FP4
peak derived from a rack figure. Through cuBLASLt from
cuBLAS~13.8.0.4, the path of \Cref{tab:setup}, B300's E2M1 GEMMs reach
$48\%$ of it at $2^{13}$ and $52\%$ at $2^{14}$.

\subsection{Pipeline Breakdown}\label{sec:steps}

The GEMMs take at least three quarters of a call at $2^{15}$, in every
configuration profiled at that size (\Cref{fig:phaseshare}), which is what
makes the GEMM count a proxy for run time (\Cref{sec:ov-channels}). At
$2^{13}$ they take $49$ to $56\%$ of it on B200: the four phases around them
scale with $n^2$ where the GEMMs scale with $n^3$, so at that size the count
alone does not order two configurations. Reduce is the largest of those four
on B200 and B300, the epilogue on the RTX~PRO~6000 at $2^{13}$ and $2^{14}$.
On B300 the INT8 call's GEMMs take $95$ to $99\%$ of
it at every size, at that GPU's reduced INT8 rate.

\subsection{Performance Against Prior Implementations}\label{sec:vsprior}

\begin{figure}[t]
\centering
\includegraphics[width=\columnwidth]{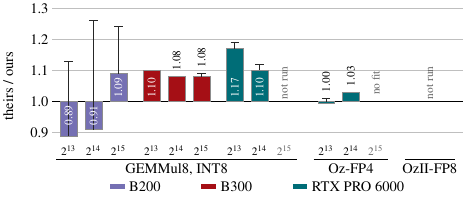}
\caption[The released implementations against ours]{Each released implementation's time over ours at the scheme it
published; above $1$ ours is faster. Ours is our CUDA library against GEMMul8
and our Triton implementation of the scheme against Oz-FP4. The two sides
ran in separate processes, GEMMul8 on the row-scaled operands of
\Cref{tab:setup} and ours on plain $\mathcal{N}(0,1)$ ones. A bar runs from
parity to the lowest ratio, and its whisker
on to the highest. GEMMul8 2.0.19 runs in accurate mode, its fastest setting
no farther from our result than native is on its operands; it sets its
scaling from its modulus product and one more INT8 product, and its result
differs from ours by $1.5\times10^{-15}$ in relative Frobenius norm.}
\label{fig:prior}
\end{figure}

Running a prior scheme in our implementation separates the scheme from the
code that runs it: the same GEMM count, a different implementation, and for
the FP4 scheme the same product. Against GEMMul8 the intervals lie above $1$
on B300 and the RTX~PRO~6000 and on B200 at $2^{15}$, and include $1$ on B200
below that size; the released FP4 code takes $1.00$ to $1.03$ times the time
of our Triton implementation of its scheme, written in the release's code
base rather than our CUDA library (\Cref{fig:prior}). Where a
configuration is compared with a prior scheme in \Cref{sec:speedup}, both
run in our library, so the difference is the
configuration's; the speedups over native FP64 and the vendor's emulation
include the implementation.

\section{Related Work}\label{sec:related}

\subsection{Residue Number Systems}\label{sec:rw-rns}

Ozaki~II is a residue number system. Channels, representatives nearest zero
and CRT reconstruction are standard there
\cite{garner_residue_1959,omondi_residue_2007}, as is multiplying integer
matrices modulo coprime word-sized moduli
\cite{doliskani_simultaneous_2018,van_der_hoeven_faster_2016}, with the words
split into limbs whose products BLAS \cite{berthomieu_multiword_2026} or INT8
Tensor Cores \cite{fan_tensorfhe_2022} form exactly, and
\Cref{sec:aroundgemms}'s estimated quotient with a correction is that
literature's approximate CRT
\cite{hung_error_1995,chervyakov_residue--binary_2017}. Here a modulus has a
cost: the alphabet decides how many GEMMs it takes, so \eqref{eq:program}
minimizes the cost of a coprime set meeting \eqref{eq:ov-crt}, where that
literature takes every modulus at one word size and maximizes the set the
width admits, by count \cite{bajard_generating_2022} or by product
\cite{gorodecky_optimal_2026}.

\subsection{The Ozaki Schemes}\label{sec:rw-schemes}

The low-precision ports of Ozaki~I \cite{ozaki_error-free_2012} keep its split
on INT8 \cite{ootomo_dgemm_2024,uchino_performance_2025}
and on FP8 \cite{mukunoki_dgemm_2025}. How many slices a call takes is then
decided per call, by the operands' range, the inner dimension and the accuracy
wanted of the result \cite{dawson_reducing_2024,brower_mixed-precision_2026},
and the GEMM count is the square of the slice count, or its triangular number
where the products of least significance are dropped
\cite{ootomo_dgemm_2024,uchino_double-precision_2026}. The
vendor's emulation estimates that count per call from the operands' exponent
span, and declines the emulation where the estimate leaves its range
\cite{schwarz_guaranteed_2026}. A CPU port to BF16 tiles fixes the count at
six slices instead, and gives up the accuracy of the scheme with it
\cite{cui_bf16_2026}.

Ozaki~II \cite{ozaki_ozaki_2025} fixes the count before any operand is seen,
which is what makes it something to minimize, and takes its moduli greedily
downward from the largest its channels admit, the rule of
\cite[\S 4.3.2]{knuth_art_1997}. The FP8 scheme
\cite{uchino_double-precision_2026} splits at base $16$, and gives a square
modulus the base $\sqrt{m}$ and the shortcut $m \mid b_m^2$ allows, which
\eqref{eq:qm} generalizes. Ozaki~2.5 \cite{matsuoka_ozaki_2026} keeps E4M3
and varies the modulus set. The FP4 scheme \cite{hayashi_dgemm_2026} fixes base $13$ and two
limbs at every channel, proves nineteen moduli minimal at that base, and
describes a partial Karatsuba set of $68$ GEMMs that its evaluation does not
use. Neither mixes one- and two-limb channels: holding every residue to a
value of the format admits $m \le 32$ at one limb on E4M3, where admissibility
(\Cref{sec:ov-cost}) requires only a member of the alphabet in each class
and admits $m \le 49$. AWE \cite{hayashi_awe_2026} extends the FP4 scheme with a
base per modulus and products of integer combinations of limbs, a class
containing \Cref{sec:ov-cost}'s forms and contained in
\Cref{sec:ov-channels}'s, and needs the $59$ GEMMs E2M1 takes in
\Cref{tab:moduli}.

Each of these derives a scheme for one alphabet. What varies here is the
alphabet, and with it the base, the limb count and the form at every channel.
Ozaki~II has also been taken to other problems, complex GEMM among them
\cite{uchino_emulation_2026}, where the moduli are the INT8 ones and the
three-product identity of \Cref{sec:ov-cost} applies to the real and imaginary
parts, as in the 3M method \cite{higham_stability_1992}.

\subsection{Implementations}\label{sec:rw-impl}

A separate line fixes the scheme and improves what surrounds the GEMMs:
GEMMul8 \cite{uchino_high-performance_2025}, kernels that fuse the residue
path into the GEMM \cite{lu_emugemm_2026}, all nineteen FP4 channels in one
launch \cite{hayashi_dgemm_2026}, an engineered conversion into residue planes
\cite{matsuoka_ozaki_2026}, and a scale-invariant fast-mode formula
\cite{kawakami_improved_2026}; \Cref{sec:vsprior} separates a lower count from
a faster kernel.

\section{Limitations}\label{sec:limitations}
\subsection{Minimizing \texorpdfstring{$Q$}{Q} Minimizes Time Only Asymptotically}

The objective of \eqref{eq:program} is the GEMM count alone. Reduce encodes
both operands at $\Theta(|\mathcal{M}|(MK + KN))$, and the epilogue and
Reconstruct make one pass over the output a modulus, at
$\Theta(|\mathcal{M}|MN)$, against the GEMMs' $\Theta(QMNK)$. The ratio of
the work outside them to the work inside is therefore
$\Theta\bigl((|\mathcal{M}|/Q)(1/M + 1/N + 1/K)\bigr)$, which vanishes only
asymptotically. That holds for a call with one accumulator over $K$. A
blocked call runs the epilogue once a slab (\Cref{sec:implslabs}), so
$K_{\mathrm{t}}$ takes the place of $K$ in the last term, and $K_{\mathrm{t}}$
is at most the configuration's $\kappa_{\mathrm{acc}}$. The epilogue then stays
at least $\Theta\bigl(|\mathcal{M}|/(Q\kappa_{\mathrm{acc}})\bigr)$ of the GEMMs
however large $K$ grows, and a configuration of higher $Q$ and larger
$\kappa_{\mathrm{acc}}$ can be the faster blocked call. Candidate configurations differ in $|\mathcal{M}|$
and in the planes they write, so they differ in the work outside the GEMMs as
well as in $Q$. Where the ratio is not small, the lower count can be the
slower call. As the size grows, the GEMMs take a larger share of the call, in
every configuration on every GPU (\Cref{fig:phaseshare}).

\subsection{Asymptotically Optimal Only When All Dimensions Grow}

Let $A$ be tall and $B$ fat, $K \ll M, N$ with $K$ fixed. The epilogue and
Reconstruct, at
$\Theta(|\mathcal{M}|MN)$, are then $\Theta\bigl(|\mathcal{M}|/(QK)\bigr)$ of
the GEMMs' $\Theta(QMNK)$, however large $M$ and $N$ grow. Let $A$ be fat and
$B$ tall, $K \gg M, N$ with $M$ and $N$ fixed. Reduce, at $\Theta(|\mathcal{M}|(MK + KN))$, is then
$\Theta\bigl(|\mathcal{M}|(1/M + 1/N)/Q\bigr)$ of them, however large $K$
grows. In neither case do the GEMMs dominate.

\subsection{Not ``True'' Emulation of DGEMM}

The configurations we time start with one lossy truncation. It keeps $53$
bits of each row's largest entry, so a much smaller entry loses its low-order
bits \cite{ozaki_error-free_2012}. They multiply the truncated operands
exactly, but they do not return $\mathrm{fl}(AB)$ for every input. Every
Ozaki~II variant truncates after the shared-exponent conversion
\cite{ozaki_ozaki_2025,uchino_high-performance_2025,uchino_double-precision_2026,uchino_error_2026,kawakami_improved_2026,hayashi_dgemm_2026},
at $53$ bits of each row's largest entry in \cite{hayashi_dgemm_2026} as here,
and in the others at a scaling set by the modulus product, so that more
moduli keep more bits.
For operands whose rows and columns span at most $2^d$, keeping $53 + d$ bits
removes the loss: the family then contains exact emulations of DGEMM, at a CRT
window $2d$ bits wider and more GEMMs (\Cref{ex:widened}). The implementation
builds both configurations of \Cref{ex:widened}, its Reduce taking operands of
$53 + d$ bits and its Reconstruct modulus products of up to $256$ bits.

\begin{example}\label{ex:widened}
At E2M1 and $K = 2^{14}$, a dynamic range of $2^{16}$ in every row of $A$ and
column of $B$ widens the CRT window by $32$ bits and raises the count from $59$
GEMMs to $78$. For such operands the widened scheme is an \emph{exact
emulation}, returning $\mathrm{fl}(AB)$. At $2^{64}$ it needs $134$.
\end{example}

\subsection{Exactness May Rely on Undocumented Accumulation}

For FP4 to FP8 on Blackwell GPUs, exactness needs
the matrix unit to add integers in $[-2^{24}, 2^{24}]$ in FP32 without
rounding. NVIDIA's ISA promises at least single precision for these additions
on the RTX~PRO~6000's \texttt{mma}, states that Hopper's \texttt{wgmma}
accumulates FP8 products above half and below single precision, and says
nothing for the \texttt{tcgen05} instructions of B200 and B300
\cite{nvidia_parallel_nodate}. Hopper's and Ada's FP8 units align the products
and accumulate them with $13$ fractional bits
\cite[\S IV-A]{khattak_accurate_2025}, about $14$ in
\cite[\S 3.3.2]{deepseek-ai_deepseek-v3_2025}; on Ada that breaks the promise
of \texttt{mma}, and on either part it would make the emulated product wrong. On B200, we estimated
$\Lambda = 2^{24}$ from the FP32 significand and checked it bitwise against a
host CRT, with every accumulator driven to $2^{24}$ at E4M3 and $K = 2^{14}$
and the products through CUTLASS. No check drove an accumulator to $2^{24}$
through cuBLASLt, or on B300 and the RTX~PRO~6000.

\subsection{May Be Less Accurate Than Native on Skewed Rows}

A native GEMM's error bound is $K u |A||B|$ to first order, with $u = 2^{-53}$
the unit roundoff \cite[\S3.5]{higham_accuracy_2002}. Let $a_i^{\max}$ be the largest magnitude in row $i$
of $A$ and $b_j^{\max}$ that in column $j$ of $B$. In ours, truncation changes
each entry by less than $2u$ times the largest in its row or column, so the
truncation term of our bound on entry $(i, j)$ is
$2u\bigl(a_i^{\max} \sum_h |b_{hj}| + b_j^{\max} \sum_h |a_{ih}|\bigr)$. If the entries of column $j$ of $B$ have about
equal magnitudes, that term alone exceeds native's error bound once $a_i^{\max}$ is
more than $(K-2)/2$ times row $i$'s mean magnitude. The
same holds with $A$ and $B$ swapped. The scheme can then be less accurate than native
\cite{ootomo_dgemm_2024,abdelfattah_analysis_2026,demmel_how_2026}. The
vendor's path sets its width per call \cite{schwarz_guaranteed_2026}. Ours
fixes its GEMM count before an operand is seen, so it cannot add GEMMs for
such a call.

\section{Conclusion}\label{sec:conclusion}

We treat the Ozaki~II schemes for low-precision formats as one family and
choose a scheme by a program that minimizes the GEMM count subject to an
exact product. We solve it exactly over candidate forms, for any format,
accumulator and inner dimension. Lower bounds over every form show that the
FP4 scheme found at $K = 2^{14}$ needs the fewest GEMMs on a stated range of
moduli. On three Blackwell GPUs the schemes run faster than native FP64 at
the sizes timed, up to $83$ times on B300. A new low-precision format needs only its values and its accumulator to
be given a scheme.

\section*{Acknowledgment}

Generative AI assistants (Anthropic Claude and OpenAI Codex) were used under
the author's direction in writing the implementation's code, searching the
literature and preparing the manuscript. The author checked the results and
takes responsibility for the content.

\printbibliography

\end{document}
\typeout{get arXiv to do 4 passes: Label(s) may have changed. Rerun}